\documentclass[10pt,letterpaper]{article}

\usepackage{cogsci}

\cogscifinalcopy 

\usepackage{booktabs}
\usepackage{amsmath}
\usepackage{graphicx} 
\usepackage{pslatex}
\usepackage[natbibapa]{apacite}
\usepackage{natbib}

\usepackage{float} 

\usepackage{xcolor}
\usepackage{hyperref}
\hypersetup{
    colorlinks,
    linkcolor={red!50!black},
    citecolor={blue!50!black},
    urlcolor={blue!80!black}
}
\usepackage{amsfonts}

\global\hyphenpenalty=1000

\title{Empathy in Explanation} 

\author{{\large \bf Katherine M. Collins\textsuperscript{1,*}} \\
    \texttt{kmc61@cam.ac.uk} \\
    \And {\large \bf Kartik Chandra\textsuperscript{2,*}} \\
    \texttt{kach@mit.edu}\\ 
    \AND {\large \bf Adrian Weller\textsuperscript{1,3}}\\
    \texttt{aw665@cam.ac.uk} \\
    \And {\large \bf Jonathan Ragan-Kelley\textsuperscript{2}} \\
    \texttt{jrk@mit.edu}
    \And {\large \bf Joshua B. Tenenbaum\textsuperscript{2}} \\
    \texttt{jbt@mit.edu} \\
    \AND
    \textsuperscript{1}University of Cambridge,
    \textsuperscript{2}MIT,
    \textsuperscript{3}The Alan Turing Institute \\
    \textsuperscript{$\ast$}These authors contributed equally to this work.
    }

\newcommand{\figureBarAlc}{
\begin{figure*}[h!]
    \centering
\includegraphics[width=0.5\linewidth]{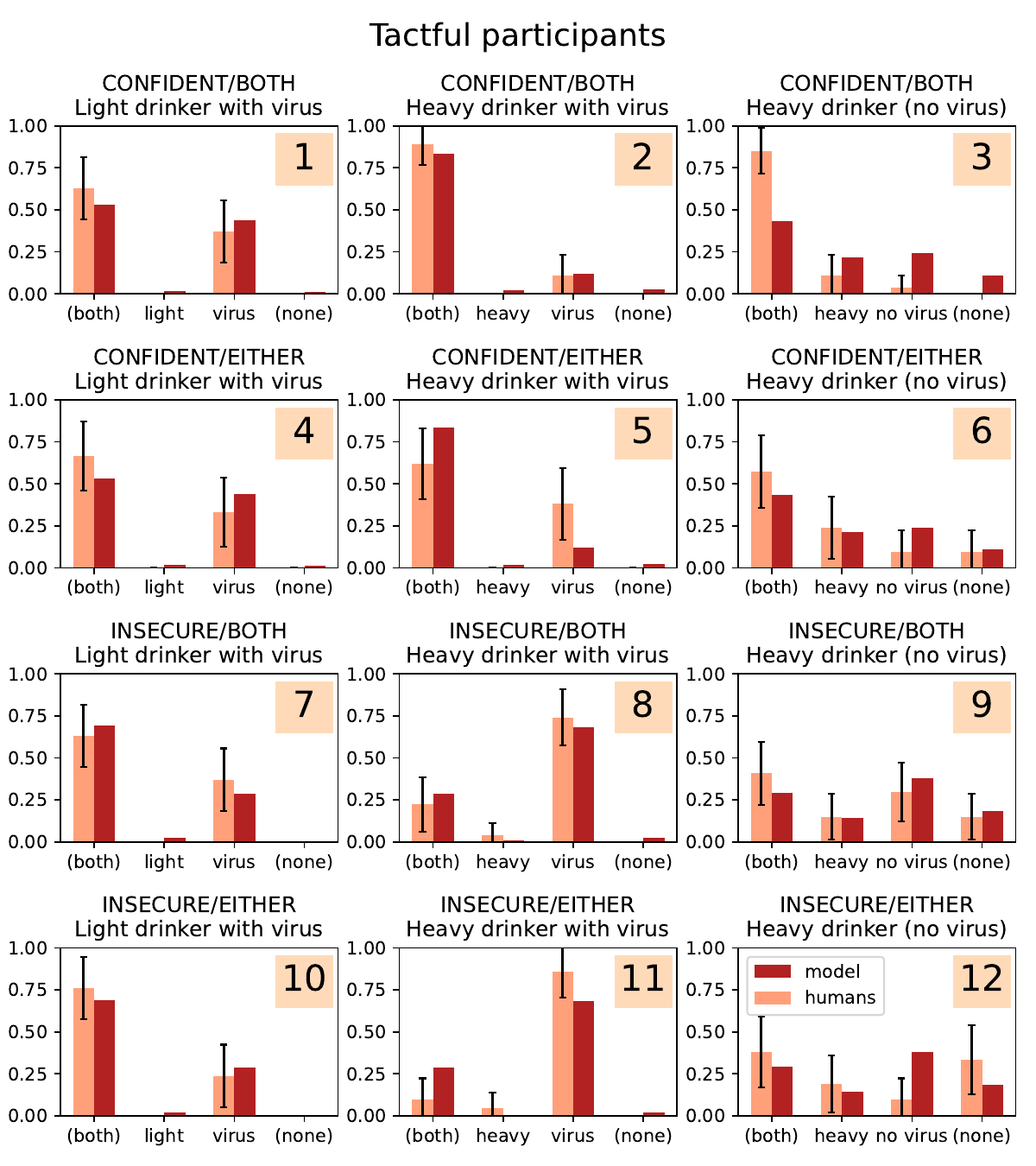}%
\includegraphics[width=0.5\linewidth]{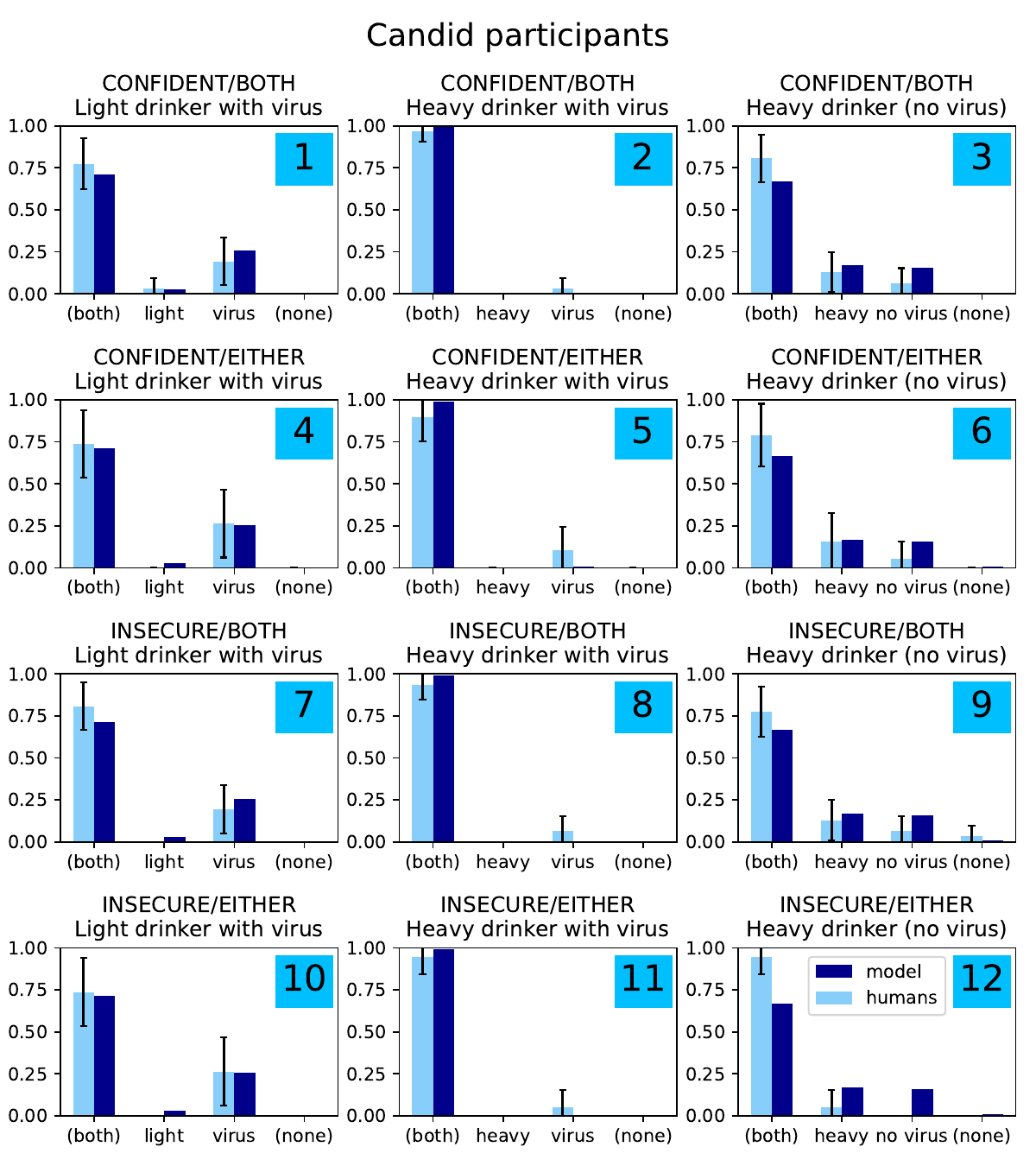}
    \caption{Human responses and model predictions across 12 scenarios. Error bars show 95\% confidence intervals across participants. The model quantitatively captures key features of human intuitions (see text in section ``Model fit'' for analysis).
    }
    \label{fig:figureBarAlc}
\end{figure*}
}

\newcommand{\figureScatterAlc}{
\begin{figure}[h!]
    \centering
\includegraphics[width=0.5\linewidth]{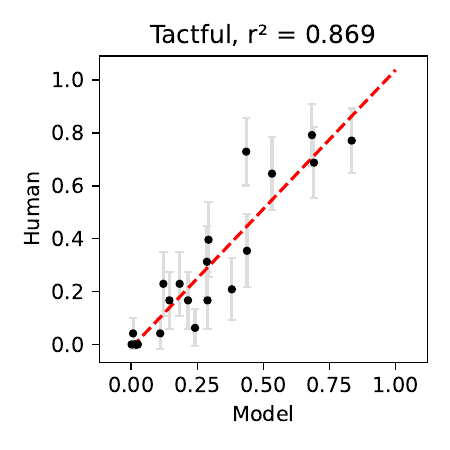}%
\includegraphics[width=0.5\linewidth]{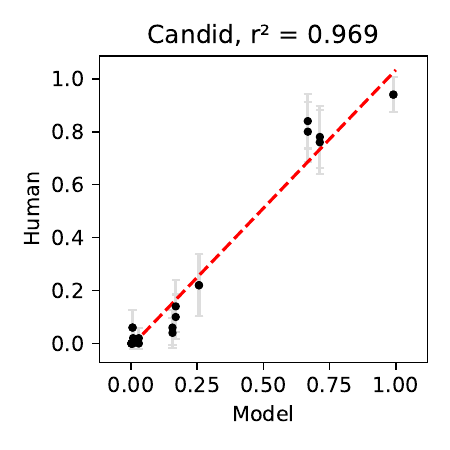}
    \caption{Correlation between humans and model. Each point represents one bar in Fig~\ref{fig:figureBarAlc}, collapsing over causal structure.}
    \label{fig:figureScatterAlc}
\end{figure}
}

\newcommand{\tableParamsAlc}{
\begin{table}[h!]
\centering
\resizebox{\columnwidth}{!}{
\begin{tabular}{lrr}
\toprule
Parameter & Tactful & Candid \\ \midrule
$\alpha_\text{social}^{R=0}$ & $0.09$ {\color{gray}$(-0.01, 2.49)$} & $0.01$ {\color{gray}$(-0.03, 0.08)$} \\
$\alpha_\text{social}^{R=1}$ & $15.12$ {\color{gray}$(10.72, 17.8)$} & $0.01$ {\color{gray}$(-0.02, 0.09)$} \\
$\log \Pr(V)$ & $-5.08$ {\color{gray}$(-5.54, -4.55)$} & $-2.7$ {\color{gray}$(-3.09, -2.34)$} \\
$\log \Pr(T)$ & $-1.99$ {\color{gray}$(-2.54, -1.38)$} & $-1.98$ {\color{gray}$(-2.47, -1.61)$} \\
$\alpha_\text{explanandum}$ & $5.22$ {\color{gray}$(2.83, 7.25)$} & $3.69$ {\color{gray}$(1.76, 6.62)$} \\
$\alpha_\text{latents}$ & $2.75$ {\color{gray}$(2.23, 3.29)$} & $6.66$ {\color{gray}$(5.63, 8.18)$} \\
\bottomrule
\end{tabular}
}
\caption{Parameter fits for our full model, presented as 95\% bootstrapped CIs computed over 100 independent samples of 80\% of our data (see text for interpretation and analysis).}
\label{tab:tableParamsAlc}
\end{table}
}
\newcommand{\tableParamsMilk}{
\begin{table}[h!]
\centering
\resizebox{\columnwidth}{!}{
\begin{tabular}{lrr}
\toprule
Parameter & Tactful & Candid \\ \midrule
$\alpha_\text{social}^{R=0}$ & $0.0$ {\color{gray}$(-0.1, 0.07)$} & $0.01$ {\color{gray}$(-0.03, 0.07)$} \\
$\alpha_\text{social}^{R=1}$ & $0.13$ {\color{gray}$(-0.02, 13.78)$} & $0.04$ {\color{gray}$(-0.01, 4.42)$} \\
$\log \Pr(V)$ & $-4.29$ {\color{gray}$(-5.2, -3.35)$} & $-2.9$ {\color{gray}$(-3.43, -2.51)$} \\
$\log \Pr(T)$ & $-0.31$ {\color{gray}$(-2.31, 0.03)$} & $-1.73$ {\color{gray}$(-2.21, -1.14)$} \\
$\alpha_\text{explanandum}$ & $6.9$ {\color{gray}$(3.36, 9.94)$} & $2.37$ {\color{gray}$(0.59, 5.31)$} \\
$\alpha_\text{latents}$ & $2.15$ {\color{gray}$(1.6, 3.05)$} & $4.91$ {\color{gray}$(4.29, 5.82)$} \\
\bottomrule
\end{tabular}
}
\caption{\emph{(Replication with milk.)} Parameter fits for full model, presented as 95\% bootstrapped CIs computed over 100 independent samples of 80\% of our data.}
\label{tab:tableParamsMilk}
\end{table}
}

\newcommand{\tableLrtAlc}{
\begin{table}[h!]
\centering
\resizebox{\columnwidth}{!}{
\begin{tabular}{lrr}
\toprule
Model class & $\text{LRT}_\text{tactful}$ & $\text{LRT}_\text{candid}$ \\ \midrule
No regret cost & $66.14$ ($**$) & $0.03$ (n.s.) \\
No inference & $80.27$ ($**$) & $184.41$ ($**$) \\
Understanding only & $94.25$ ($**$) & $187.28$ ($**$) \\
No understanding & $14.72$ ($**$) & $13.42$ ($**$) \\
\bottomrule
\end{tabular}
}
\caption{Log likelihood ratio test values for the ablations versus full model $**$ indicates p-value $< 0.01$.}
\label{tab:tableLrtAlc}
\end{table}
}

\newcommand{\tableLrtMilk}{
\begin{table}[]
\centering
\resizebox{\columnwidth}{!}{
\begin{tabular}{lrr}
\toprule
Model class & $\text{LRT}_\text{tactful}$ & $\text{LRT}_\text{candid}$ \\ \midrule
No regret cost & $22.84$ ($**$) & $4.67$ (n.s.) \\
No inference & $49.14$ ($**$) & $135.1$ ($**$) \\
Understanding only & $49.27$ ($**$) & $137.68$ ($**$) \\
No understanding & $65.39$ ($**$) & $2.45$ (n.s) \\
\bottomrule
\end{tabular}
}
\caption{\emph{(Replication with milk.)} Log likelihood ratio test values for ablations versus full model. $**$ indicates $p<0.01$.}
\label{tab:tableLrtMilk}
\end{table}
}

\newcommand{\figureScatterMilk}{
\begin{figure}[h!]
    \centering
\includegraphics[width=0.5\linewidth]{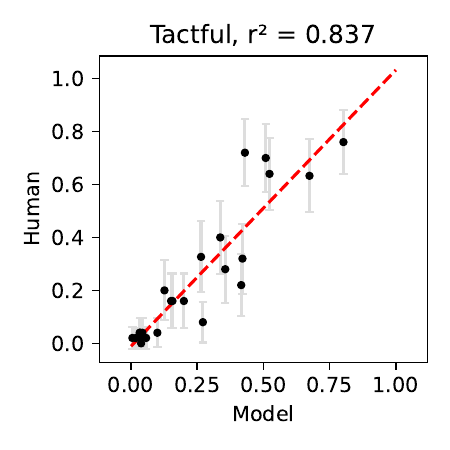}%
\includegraphics[width=0.5\linewidth]{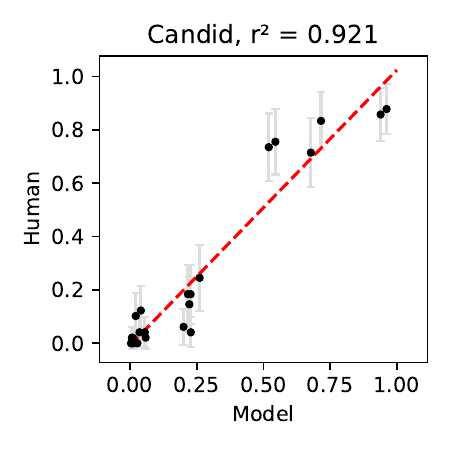}
    \caption{\emph{(Replication with milk.)} Correlation between human responses and model predictions.}
    \label{fig:figureScatterMilk}
\end{figure}
}

\newcommand{\tableRAlc}{
\begin{table}[h!]
\centering
\resizebox{\columnwidth}{!}{
\begin{tabular}{lrr}
\toprule
Model Class & $r^2~\text{tactful}$ & $r^2~\text{candid}$ \\ \midrule
Full model & \textbf{0.84} {\color{gray}$(0.76, 0.91)$}  & \textbf{0.96} {\color{gray}$(0.93, 0.99)$} \\
No regret cost & $0.58$ {\color{gray}$(0.51, 0.65)$} & \textbf{0.96} {\color{gray}$(0.92, 0.99)$} \\
No inference & $0.62$ {\color{gray}$(0.55, 0.69)$} & $0.53$ {\color{gray}$(0.50, 0.56)$} \\
Understanding only & $0.56$ {\color{gray}$(0.52, 0.65)$} & $0.52$ {\color{gray}$(0.48, 0.55)$} \\
No understanding & $0.76$ {\color{gray}$(0.45, 0.86)$} & $0.95$ {\color{gray}$(0.92, 0.98)$} \\
\bottomrule
\end{tabular}
}
\caption{Correlations between human data and competing models, presented as 95\% bootstrapped CIs computed over 100 independent samples of 80\% of our data.}
\label{tab:tableRAlc}
\end{table}
}

\newcommand{\tableRMilk}{
\begin{table}[h!]
\centering
\resizebox{\columnwidth}{!}{
\begin{tabular}{lrr}
\toprule
Model Class & $r^2~\text{tactful}$ & $r^2~\text{candid}$ \\ \midrule
Full model & \textbf{0.76} {\color{gray}$(0.65, 0.87)$} & \textbf{0.90} {\color{gray}$(0.86, 0.94)$}\\
No regret cost & $0.70$ {\color{gray}$(0.61, 0.78)$} & \textbf{0.90} {\color{gray}$(0.85, 0.94)$} \\
No inference & $0.65$ {\color{gray}$(0.57, 0.73)$} & $0.54$ {\color{gray}$(0.51, 0.58)$} \\
Understanding only & $0.65$ {\color{gray}$(0.60, 0.72)$} & $0.54$ {\color{gray}$(0.48, 0.59)$} \\
No understanding & $0.61$ {\color{gray}$(0.52, 0.85)$} & $0.90$ {\color{gray}$(0.84, 0.95)$} \\
\bottomrule
\end{tabular}
}
\caption{\emph{(Replication with milk.)} Correlations between human data and competing models, presented as 95\% bootstrapped CIs computed over 100 independent samples of 80\% of our data.}
\label{tab:tableRMilk}
\end{table}
}

\begin{document}

\maketitle

\begin{abstract}
Why do we give the explanations we do? Recent work has suggested that we should think of explanation as a kind of cooperative social interaction, between a why-question-asker and an explainer. Here, we apply this perspective to consider the role that emotion plays in this social interaction. We develop a computational framework for modeling explainers who consider the emotional impact an explanation might have on a listener. We test our framework by using it to model human intuitions about how a doctor might explain to a patient why they have a disease, taking into account the patient's propensity for regret. Our model predicts human intuitions well, better than emotion-agnostic ablations, suggesting that people do indeed reason about emotion when giving explanations.

\end{abstract}

\section{Introduction}

Why do we answer ``why'' questions the way we do? A long line of work on explanation \citep[see][for reviews]{lombrozo2006structure, lombrozo2012explanation, miller2019explanation} has shown that good explanations are contrastive \citep{lipton1990contrastive, riveiro2021thats}, selective \citep{lombrozo2007simplicity, gerstenberg2020expectations, poesia2022left}, and causal \citep{josephson1996abductive, hilton1996mental, mcclure2002goal} in nature.

However, recent work has argued that explanations are also \emph{social} in nature: our intuitions about explanation are contextual, depending on the mental states of both the explainer and the listener~\citep{harding2025communication}. For example, \citet{kirfel2024explain} show that explanations are sensitive to the explainer's assessment of the listener's \emph{goals:} good explanations help the listener make good interventions. Similarly, \citet{chandra2024cooperative} show that explanations are sensitive to the explainer's assessment of the listener's \emph{beliefs:} good explanations infer and address the misconceptions that led to the ``why?'' question in the first place.

In this paper, we extend this argument by studying another type of mental state that may be relevant to the social dynamics of explanation: \emph{emotions}. Explanations can create a wide variety of affective reactions, because many emotion concepts are closely related to notions of causality. For example, an explanation that attributes causal responsibility to a listener's actions can cause that listener to feel regret, guilt, or pride---while an alternate explanation that absolves the listener from responsibility might cause that listener to feel relief instead.
Hence, if explanations are indeed subject to the principles of cooperative social interaction, we should expect cooperative explainers to take the listener's emotional response into account when crafting an explanation.

In this paper, we test this prediction by studying human intuitions about explanation in an emotionally-charged doctor-patient communication setting. In our experiment, a patient with a terminal disease asks their doctor what caused the disease. Participants in our experiment are tasked with playing the role of the doctor and giving an explanation. Specifically, they choose whether to communicate that the disease was caused by the patient's lifestyle choices (e.g.\ excessive drinking), an external factor beyond their control (e.g.\ a virus), or both. We asked whether participants would expect the doctor to adjust their explanation to minimize the extent to which the patient \emph{regrets} their past actions.

We modeled participants' judgments on this task using a variety of computational models, which instantiate competing hypotheses about what role (if any) emotion may play in explanation. We built on \citet{chandra2024cooperative}'s framework for modeling explanation as a type of rational communication that follows Gricean maxims \citep{grice1975logic, sumers2023reconciling, frank2012predicting, goodman2016pragmatic}, extending their framework to account for the emotional impact an explanation might have on the listener. Specifically, we introduced a new emotion-sensitive term in the explainer's utility function, which is grounded in recent work on Bayesian models of emotion appraisal \citep{houlihan2023emotion} and intervention \citep{chen2024intervening}.

In the population we studied, we found an even split between participants who considered the patient's regret when giving explanations, and participants who preferred to be forthcoming with all information. Our regret-sensitive model predicted both subgroups' intuitions well ($r^2=0.87$ and $r^2=0.97$, respectively). Critically, for the former subgroup, our model was more predictive than an alternate model that ablated the regret term ($r^2=0.58$), suggesting that some people indeed take emotion into consideration when giving explanations.

\section{Review of computational framework}

Our computational framework extends \citeauthor{chandra2024cooperative}'s framework of explanation as rational communication. Under this framework, an explainer (Alice) can be modeled as \emph{inferring and intervening on} the question-asker's (Bob's) mental model of the situation at hand. Concretely, this framework assumes that the question-asker asks a ``why?'' question when faced with an event $e$ that leads to an expectation violation under Bob's current mental model of the world, i.e.\ when $Pr(e \mid m_\text{asker}) \ll 1$ under mental model $m_\text{asker}$. The explainer infers $m_\text{asker}$ from the question asked, and then chooses an explanation $u$ (``utterance'') among a space of candidate explanations by softmax over Alice's utility $V$: that is, $\Pr(u) \propto \exp(\beta \cdot V(u))$. The explainer's utility $V$ is decomposed into two key terms:
\begin{enumerate}
    \item An \emph{understanding term}, which measures the degree to which the question-asker's expectation violation would be resolved by updating Bob's mental model according to the information in $u$, i.e.\  $\Pr(e \mid \text{update}(m_\text{asker}, u))$. If this probability is high, then Bob's
    expectation-violation has been resolved.
    \item A \emph{social term}, which measures additional social considerations that may be in play during an explanation-interaction.
\end{enumerate}
For the social term, \citeauthor{chandra2024cooperative} specifically considered the phenomenon of ``mansplaining'': offering redundant or obvious information in an explanation could embarrass the question-asker by implying that the explainer Alice thinks that the question-asker Bob is ignorant. Here, we generalize the social term to more broadly consider the role of the question-asker Bob's affective response to the explainer Alice's explanation.
We model possible emotional costs in terms of \emph{appraisal variables} \citep{moors2014flavors, frijda1986emotions, houlihan2023emotion, lazarus1991emotion}, which are computed with respect to the question-asker Bob's updated mental model upon hearing the explanation $u$.

In the next section, we show how this framework can be applied in the setting of our experiment, in order to model how explainers take the emotion of \emph{regret} into account when giving explanations.

\section{Experiment}

\subsection{Setting}

\begin{figure}
    \centering
    \includegraphics[width=\linewidth]{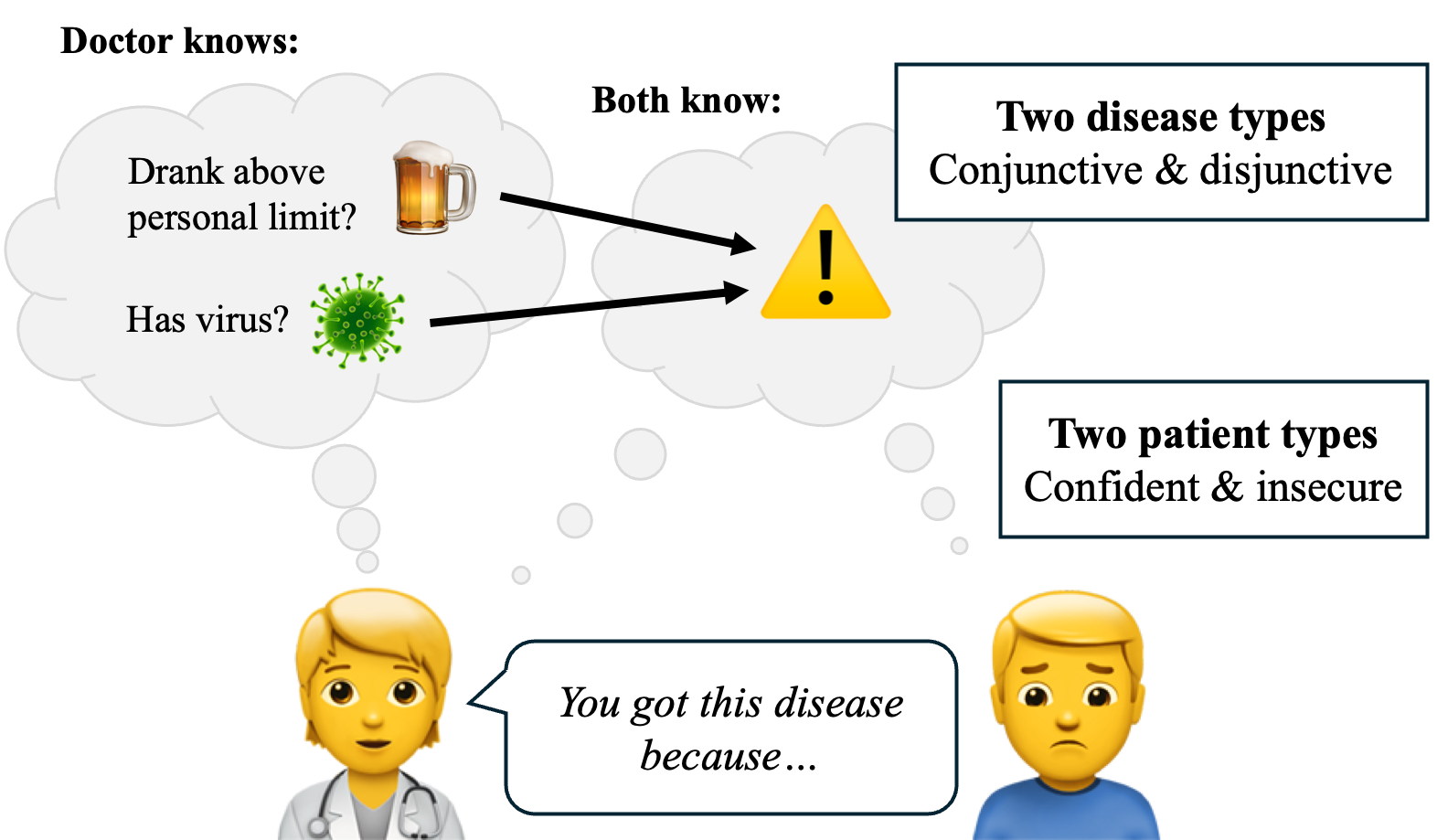}
    \caption{Schematic representation of our experimental paradigm. A disease is caused by two possible factors. The doctor knows the patient's test results and temperament, and decides what information to communicate in the explanation.}
    \label{fig:scenarios}
\end{figure}

To study the role of emotion in explanation, we considered a domain where both informativity and emotional response are at stake in explanation: medical communication. In our experiments, a doctor explains to a patient why they have a disease (Figure~\ref{fig:scenarios}). We considered fatal diseases with two potential causes:
\begin{itemize}
    \item A \textbf{virus}, which patients have no control over contracting, and whose presence can only be detected by a blood test conducted by a doctor.
    \item Excessive \textbf{alcohol consumption} in youth, which patients \emph{did} have agency over. The threshold for ``excessive'' varies from person to person and can only be calculated by a doctor based on the patient's medical profile, so patients do not know a priori whether they drank excessively.
\end{itemize}

Following prior work in explanation \citep{kominsky2015causal, icard2017normality}, we considered two types of diseases with different causal structures: \emph{``conjunctive''}, where \emph{both} factors are needed for a patient to get the disease, and \emph{``disjunctive''}, where \emph{either} factor alone can cause the disease.
The doctor and patient both understand the causal structure of the disease. However, only the doctor knows whether the patient has the virus, and whether the patient's drinking was excessive. The doctor thus explains to the patient why they have the disease by mentioning either, both, or (rarely) neither of the two causes in the explanation.

Finally, we considered two types of patients: those who are temperamentally \textbf{confident} about past decisions (``comfortable with past choices and calmly accepting of consequences'') and \textbf{insecure} about past decisions (``often wonders whether drinking in college was the greatest mistake of their life''). Participants were told that the doctor knows the patient's temperament based on past medical records.

Because future interventions (as studied by \citeauthor{kirfel2024explain}) were beyond the scope of this experiment, we emphasized to participants that the doctor's explanation could not influence the patient's future actions. We emphasized this in two ways: first, we told participants that the disease was fatal and that patients only had a few days left to live, and second, we told participants that the disease was linked to \emph{past} drinking, and that all patients had quit alcohol for good after college.

To summarize our experimental design: we considered $2\times2\times3=12$ scenarios, varying the type of disease (conjunctive/disjunctive), the type of patient (confident/insecure), and the test results received by the doctor (has virus, drank excessively, or both). This design is summarized in Figure~\ref{fig:scenarios}.

\subsection{Procedure}
We recruited 125 participants from Prolific~\citep{palan2018prolific} in an IRB-approved study. Participants were based in the United States and were required to be fluent in English. Participants took on average of 15 minutes to complete the experiment and were paid at a rate of \$15 per hour.
Each participant was randomly assigned one causal structure (conjunctive/disjunctive), and completed a quiz to confirm that they understood the structure. We automatically discarded responses from 25 participants who failed the quiz.

Participants were then shown six scenarios (crossing patient temperament and true cause(s) of the patient's disease), and tasked with playing the role of the doctor: explaining to a patient why they have the disease by choosing whether or not to mention each of the two possible causes. 

At the end of the experiment, we asked participants directly whether they took the patients' temperament (confident/insecure) into account when giving explanations. We expected some people to say that they did, to avoid needless cruelty towards insecure patients, while others to say they did not, on the principle that doctors should always be fully forthcoming about all relevant information \citep{faden1981disclosure}. We used this self-reported information to code participants as \textbf{tactful} (sensitive to the patient's temperament) or \textbf{candid} (not sensitive to the patient's temperament).  We manually inspected all participants' textual justifications of their responses and found that participants' self-reports were generally consistent with their justifications, with the exception of 3 of the 125 participants. We manually corrected the classifications of these three participants for the remainder of our analyses.

\subsection{Computational model}

We applied the computational framework discussed above to model participants' intuitions for how the doctor might take the patient's regret into account when giving explanations.

Under our model, the patient knows how much they drank in their youth ($D$). However, the patient has uncertainty over two binary variables: their personal safe threshold for drinking ($T \in \{\text{low}, \text{high}\}$), and whether or not they have the virus ($V \in \{\text{has virus}, \text{does not have virus}\}$). This uncertainty is given by their priors $\Pr(T)$ and $\Pr(V)$, which are free parameters to the model. The patient understands how these possible causes affect the probability of getting sick, i.e.\ they can compute $\Pr(S \mid D, T, V)$. The patient then observes that they are sick with the disease ($S$), and asks the doctor ``why?''

Like the patient, the doctor knows how much the patient drank in their youth ($D$). In addition, the doctor calculates $T$ and measures $V$ with a blood test. Finally, the doctor knows the patient's temperament ($R$): whether they are confident ($R=0$) or insecure about their past choices ($R=1$).

To answer the patient's ``why?'' question, the doctor chooses among four possible utterances: $U=\{(T, V), (T, \phi), (\phi, V), (\phi, \phi)\}$, where $T$ indicates communicating the patient's threshold (effectively, whether or not they drank too much), $V$ indicates communicating whether or not the patient has the virus, and $\phi$ represents not mentioning a given fact. (In other words, the doctor can say any subset of the two possible true statements, including the empty subset.)

\figureBarAlc

The doctor selects $u\in U$ by softmax over their total utility, \( \Pr(u) \propto \exp(\alpha \cdot \mathcal{V}(u)) \).
The total utility is made up of three terms, $\mathcal{V}_\text{explanandum}$, $\mathcal{V}_\text{latents}$, and $c_\text{social}$, which are combined linearly:
\begin{align}
\mathcal{V}(u) &= \alpha_\text{explanandum} \cdot \mathcal{V}_\text{explanandum}(u) \\
&+ \alpha_\text{latents} \cdot \mathcal{V}_\text{latents}(u) \\
&- \alpha_\text{social}^R \cdot c_\text{social}(u).
\end{align}
The understanding portion of the utility measures the patient's understanding of the explanandum, conditioned on the doctor's utterance $u$. We calculate this in two parts: the likelihood the patient now assigns to getting sick (given by $\mathcal{V}_\text{explanandum}(u)=\Pr(S \mid u)$), and the likelihood the patient now assigns to the latent factors $T, V$ conditioned on the doctor's utterance $u$ (given by $\mathcal{V}_\text{latents}(u)=\Pr(T, V \mid u)$). The latter is important because even though the doctor is constrained to make true statements, they might additionally wish to avoid creating a false \emph{implication} through their explanation (i.e.\ ``lying by omission''). For example, if the patient had both factors and the doctor only mentioned one, then by the principle of ``explaining away,'' the patient may reach a false conclusion about the presence of the other factor.

Finally, we model the social cost of potential patient regret by computing the ``counterfactual utility'' appraisal term defined by \citet{houlihan2023emotion}, given in this case by $c_\text{social}(u) = \Pr(S=1 \mid \text{do}(D=0), u)$: the counterfactual likelihood of getting sick if they had abstained from alcohol ($D=0$), taking into account what the patient learns from the explanation ($u$). We separately fit weights $\alpha_\text{social}^{R=0}$ for confident patients and $\alpha_\text{social}^{R=1}$ for insecure patients.

To summarize, the $6$ free parameters to our model are: the patient's priors over the latents $\Pr(T)$ and $\Pr(V)$, the doctor's rewards for the patient's understanding $\alpha_\text{explanandum}$ and $\alpha_\text{latents}$, and the doctor's regret cost $\alpha_\text{social}^{R=0}$ and $\alpha_\text{social}^{R=1}$ for confident and insecure patients, respectively.
We fit our parameters to behavioral data by gradient descent, using the \textit{memo} probabilistic programming language \citep{chandra2025memo}, and we applied an L1 regularizer with $\lambda=0.005$ to prevent overfitting.

\subsection{Alternate models}

We compared our full model, described above, to the following alternate models.

\begin{enumerate}
    \item No regret cost, $\alpha_\text{social}^{\{0,1\}} = 0$. This model represents an explainer who does not weigh the regret experienced by the patient in response to an explanation.
    \item No inference cost, $\alpha_\text{latents}=0$. This model represents an explainer who does not weigh whether the patient comes to an incorrect conclusion about a latent variable.
    \item No regret \emph{or} inference cost, $\alpha_\text{social}^R = \alpha_\text{latents} = 0$. This model represents an explainer who only tries to minimize the listener's post-explanation surprisal, without regard to possible incorrect inferences about latent variables.
    \item No understanding reward, $\alpha_\text{explanandum} = 0$. This model represents an explainer who does not weigh the listener's post-explanation surprisal.
\end{enumerate}
We expected that our full model would predict participants' behavior better than ablations 1-4, especially among tactful participants, suggesting that people weigh both understanding and emotion when giving explanations.


\figureScatterAlc

\subsection{Results}

Recall that we coded participants as being either tactful or candid based on their self-reports. We found that participants were approximately evenly split among these two categories (49\% and 51\%, respectively). Our data, segmented by participant type, is shown in Figure \ref{fig:figureBarAlc}.

These two kinds of participants differed meaningfully in their behavior. We saw a significant difference ($p < .001$) in how often the different sets of participants communicate both $T$ and $V$ to patients: tactful participants communicate both factors only $56.6\%$ of the time, while candid participants do so $84.3\%$. As expected, tactful participants were more reluctant to communicate about $T$ when a patient was insecure than candid ones were (see scenarios \#5, \#6, \#11, \#12).

We found that participants did not significantly differ in their behavior on conjunctive vs.\ disjunctive causal structures. We believe this is because participants did not fully understand the nuances of the causal structures as we presented them (as evidenced also by the high rate of failing the quiz, $20\%$). Thus, for the remainder of our analyses, we collapsed data across the two causal structures. In our models, we set the probability of disease proportional to the number of factors present: epsilon if neither factor is present, $0.25$ if only one factor is present, and $0.5$ if both factors are present.

\paragraph{Model fit}  We fit the 6 free parameters of our model, separately for both tactful and candid participants (Figure~\ref{fig:figureScatterAlc}). We find that our model captures both types of participants' judgments well ($R^2 = 0.869$ and $R^2=0.969$, respectively).

For example, when communicating with an insecure heavy drinker with the virus (scenarios \#8/\#11), tactful participants (left/red) only mention the virus, presumably to prevent the patient from regretting past drinking. Candid participants (right/blue), however, mention both factors. Tactful participants \emph{do} mention both factors if the patient is confident and unlikely to feel regret (scenarios \#2/\#5), or a light drinker who has nothing to regret in the first place (scenarios \#7/\#10). Our model quantitatively captures all of these intuitions.

\tableParamsAlc
\tableLrtAlc
\tableRAlc

\paragraph{Analysis of parameter fits} Table~\ref{tab:tableParamsAlc} shows our full model's parameter fits. We can interpret these fitted parameters as follows:
\begin{enumerate}
    \item For tactful participants, $\alpha_\text{social}^{R=1}$ is significantly higher than 0, while $\alpha_\text{social}^{R=0}$ is not. This means that tactful participants weigh the regret cost significantly for insecure patients, but not at all for confident patients.
    
    However, for candid participants, neither $\alpha_\text{social}^{R=0}$ nor $\alpha_\text{social}^{R=1}$ are significantly different from 0. This means that candid participants do not weigh the regret cost for patients of either temperament.
    \item For both kinds of participants, the priors fit to small probabilities, and $\Pr(V) \ll \Pr(T)$. In other words, participants expected patients to have low priors for either factor, and considered the rare virus to be less likely \emph{a priori} than excessive drinking.
    \item For both kinds of participants, $\alpha_\text{explanandum}$ is positive (and of similar magnitude). This means that both kinds of participants sought to reduce the patient's expectation violation.
    \item For both kinds of participants, $\alpha_\text{latents}$ is positive, but it is \emph{lower} for tactful participants and \emph{higher} for candid participants. In other words, tactful participants incur a much lower cost of creating a false inference about latent variables.
\end{enumerate}

\paragraph{Model comparison}

Our full model captured human intuitions better than all four of our ablations, as measured by both a Likelihood Ratio Test (Table~\ref{tab:tableLrtAlc}) and by comparison of bootstrap confidence intervals over $r^2$ values (Table~\ref{tab:tableRAlc}). An important exception is that our model can predict the behavior of \emph{candid} participants even when the regret terms are ablated. This is to be expected: candid participants are those who self-reported not taking regret into account. Together, these results support the hypothesis that some people do indeed take both emotion and understanding into account when giving explanations (at least, when they consider it appropriate to do so).


\paragraph{Replication with milk}

To test the generality of our findings, we re-ran our experiment above with $N=120$ participants, replacing references to alcohol with references to milk. We reasoned that patients may be likelier to regret drinking alcohol than consuming milk.

Our full model (re-fit to new data) continued to predict human responses well (Figure~\ref{fig:figureScatterMilk}). Most of our above findings related to ablations and parameter fits replicated straightforwardly (Tables~\ref{tab:tableLrtMilk}, \ref{tab:tableRMilk}, and \ref{tab:tableParamsMilk}). There were two key deviations:
\begin{enumerate}
    \item For tactful participants, $\alpha_\text{social}^{R=1}$ was not quite significantly higher than 0, suggesting that participants weighed a patient's regret about consuming milk less strongly than a patient's regret about alcohol. 

    \item For candid participants, ablating the $\alpha_\text{explanandum}$ did not significantly reduce the model fit (by either LRT or comparison of $r^2$ values). This suggests that in this experiment, candid participants were even more concerned with ensuring that the patients understood their true condition.
\end{enumerate}



\newpage

\section{Discussion}

If explanations are indeed fundamentally social in nature, then we should expect them to be sensitive not only to the explainer and listener's beliefs and desires, but also more sophisticated mental states, such as emotions. In this paper, we presented a computational framework for modeling the role emotion plays in explanation. In our framework, the explainer's utility includes a new term for the emotional impact on the listener. Using this framework, we modeled human behavior in a doctor-patient communication setting, showing that people do indeed weigh both understanding and emotion in explanation. Our work connects to a growing body of work on explanation as a social interaction. More broadly, it contributes to a wider literature on how people balance referential and social utilities when communicating: for example, in the case of polite speech \citep{yoon2020polite}.

Our computational framework can be extended in many different directions. For example, we focused here on the emotion of regret; however, many other emotions (e.g.\ guilt) are connected to notions of causality. To integrate other emotions into our models, we would introduce new terms into the explainer's social utility, which would be computed in terms of those emotions' respective appraisal variables.

Our model also does not yet take into account higher-level pragmatic reasoning on the part of either the question-asker or the patient. For example, if a doctor seems to be suspiciously avoiding any mention of alcohol, a patient might think that the doctor is withholding information, and conclude that the disease might have been their fault after all. Hence, in some cases a doctor may make a point to mention to the patient that the disease was \emph{not} caused by excessive alcohol consumption, even if the patient had low priors on that hypothesis. This reasoning is conceptually straightforward to introduce into our models: we can model a ``level-2'' patient who reasons about a level-1 doctor as described in this paper, and then consider a level-2 doctor who reasons about the level-2 patient~\citep{camerer2004cognitive}.

Here, we considered a cooperative explainer who seeks to prevent negative emotions---but explanation-interactions need not always be cooperative. For example, during a trial, a prosecutor cross-examining a witness may adversarially challenge that witness with uncomfortable ``why?'' questions, designed such that the witness' discomfort in answering creates the impression of guilt to a juror. To model such sophisticated scenarios, we might consider jurors who \emph{interpret} explanations by reasoning about emotion.

Finally, taking a step back, we hope that these findings can not only help us develop our theoretical understanding of explanation, but can also be applied in practice to build more human-compatible AI thought partners \citep{collins2024building}: for example, to build AI teaching assistants that take emotions like shame and embarrassment into account when giving explanations in response to students' questions \citep{chandra2024watchat}.

\section*{Acknowledgments}

We thank Dae Houlihan, Desmond Ong, Lio Wong, Tyler Brooke-Wilson, Tony Chen, and Danny Collins for valuable conversations that informed this work. KMC acknowledges funding from the Cambridge Trust and King's College Cambridge. KC was supported by the Hertz Foundation and the NSF GRFP. AW  acknowledges  support  from  a  Turing  AI  Fellowship  under grant  EP/V025279/1 and the Leverhulme Trust via CFI. This work is supported (in part) by ELSA - European Lighthouse on Secure and Safe AI funded by the European Union under grant agreement No. 101070617. Views and opinions expressed are however those of the author(s) only and do not necessarily reflect those of the European Union or European Commission. 

\bibliographystyle{apacite}

\setlength{\bibleftmargin}{.125in}
\setlength{\bibindent}{-\bibleftmargin}

\bibliography{cogsci24, references}

\clearpage

\appendix


\figureScatterMilk
\tableLrtMilk
\tableRMilk
\tableParamsMilk

\end{document}